\documentclass[titlepage]{article}
\usepackage[]{graphicx}
\usepackage[]{color}

\usepackage{amsmath,amsfonts,amssymb,amsthm,epsfig,epstopdf,titling,url,array}


\usepackage[section]{placeins}
\usepackage{setspace}
\theoremstyle{plain}
\newtheorem{thm}{Theorem}[section]
\newtheorem{lem}[thm]{Lemma}

\theoremstyle{definition}

\usepackage{placeins}

\theoremstyle{remark}

\makeatletter
\def\maxwidth{ %
  \ifdim\Gin@nat@width>\linewidth
    \linewidth
  \else
    \Gin@nat@width
  \fi
}
\makeatother

\definecolor{fgcolor}{rgb}{0.345, 0.345, 0.345}

\usepackage{framed}
\makeatletter
 {\par\unskip\endMakeFramed%
 \at@end@of@kframe}
\makeatother

\definecolor{shadecolor}{rgb}{.97, .97, .97}
\definecolor{messagecolor}{rgb}{0, 0, 0}
\definecolor{warningcolor}{rgb}{1, 0, 1}
\definecolor{errorcolor}{rgb}{1, 0, 0}

\usepackage{alltt}
\usepackage{enumitem}
\usepackage{amsthm}
\usepackage{amsmath}
\usepackage{amssymb}
\usepackage{amsfonts}

\usepackage[style=authoryear,maxcitenames=2, doi=false]{biblatex}

\bibliography{blipVar}

\usepackage[letterpaper, portrait, lmargin=1in, rmargin=1in,
bmargin = 1.35in, tmargin = 1.35in]{geometry}
\usepackage[english]{babel}
\usepackage{graphicx}
\usepackage{float}

\usepackage{caption}

\IfFileExists{upquote.sty}{\usepackage{upquote}}{}

\title{Kernel Smoothing of the Treatment Effect CDF}
\author{Jonathan Levy, Mark van der Laan}
\begin{document}

\begin{titlepage}

\maketitle
\begin{abstract}
The strata-specific treatment effect or so-called blip for a randomly drawn strata of confounders defines a random variable and a corresponding cumulative distribution function.  However, the CDF is not pathwise differentiable, necessitating a kernel smoothing approach to estimate it at a given point or perhaps many points.  Assuming the CDF is continuous, we derive the efficient influence curve of the kernel smoothed version of the blip CDF and a CV-TMLE estimator.   The estimator is asymptotically efficient under two conditions, one of which involves a second order remainder term which, in this case, shows us that knowledge of the treatment mechanism does not guarantee a consistent estimate. The remainder term also teaches us exactly how well we need to estimate the nuisance parameters (outcome model and treatment mechanism) to guarantee asymptotic efficiency.  Through simulations we verify theoretical properties of the estimator and show the importance of machine learning over conventional regression approaches to fitting the nuisance parameters. We also derive the bias and variance of the estimator, the orders of which are analogous to a kernel density estimator. This estimator opens up the possibility of developing methodology for optimal choice of the kernel and bandwidth to form confidence bounds for the CDF itself.   
\end{abstract}
\end{titlepage}

\section{background and Motivation}
The stratum-specific treatment effect or blip \parencite{gillrobins2001} function is defined as random variable given by the average treatment effect for a randomly drawn stratum of confounders.  Estimating the cumulative distribution function or CDF of the blip function therefore estimates the proportion of the population that has an average treatment effect at or below a given level.  Because clinicians treat patients based on confounders, such is of interest as to evaluating treatment.  However, we will see the blip CDF is not pathwise differentiable so we need to estimate a kernel-smoothed version, which can be of interest in and of itself.  Such might also provide a pathway to forming confidence intervals for the blip CDF.  \\

Much consideration has been given to the distribution of $Y_1-Y_0$, where $Y_a$ is the counterfactual outcome under the intervention to set treatment to $a$, as per the neyman-rubin potential outcomes framework  \parencite{neyman1923, rubin1974}.  Neyman, 1923, realized that even in estimating the mean of $Y_1-Y_0$, the impossibility of identifying the correlation of $Y_1, Y_0$ hampered variance estimation in small samples.  Assumptions needed to estimate the joint distribution of $Y_1$ and $Y_0$ are hard to verify.  Cox, 1958 assumes a constant treatment effect for pre-defined subgroups, while Fisher, 1951 suggests one can essentially view the counterfactual $Y_1-Y_0$ by careful design.  Heckman and Smith, 1998 estimate the quantiles of $Y_1-Y_0$ via the assumption of quantiles being preserved from $Y_1$ to $Y_0$ given a strata of confounders.  Without strong assumptions, using tail bounds \parencite{frechet1951} to estimate the quantiles of $Y_1-Y_0$ via the marginals of $Y_1$ and $Y_0$ tends to leave too big of a measure of uncertainty to be useful \parencite{heckman}.  Heckman also mentions that his analysis becomes much easier if $Y_1-Y_0$ remains fixed for a given stratum, i.e., $Y_1-Y_0 = E[Y_1-Y_0 \mid W]$, for which we aim to estimate the CDF.  \nocite{heckman1998}

\section{Data}
Our full data, including unobserved measures, is assumed to be generated according to the following structural equations \parencite{Wright, Strotz, Pearl:2000aa}.  We can assume a joint distribution, $U=(U_{W},U_{A},U_{Y})\sim P_{U}$, an unknown distribution of unmeasured variables. $X = (W,A,Y)$ are the measured variables.  In the time ordering of occurrence we have $W=f_{W}(U_{W})$ where $W$ is a vector of confounders, $A=f_{A}(U_{A},W)$, where $A$ is a binary treatment and $Y=f_{Y}(U_{Y},W,A)$, where $Y$ is the outcome, either binary or bounded continuous.  We thusly define a distribution $P_{U,X}$, via $(U,X) \sim P_{U,X}$.\\

The full-data model, $\mathcal{M}^{F}$, which is non-parametric, consists of all possible $P_{UX}$. The observed data model, $\mathcal{M}$, is linked to $\mathcal{M}^{F}$ in that we observe $O=(W,A,Y)\sim P$ where $O=(W,A,Y)$ is generated by $P_{UX}$ according to the structural equations above.  Our true observed data distribution, $P_{0}$, is therefore an element of a non-parametric observed data model, $\mathcal{M}$.  In the case of a randomized trial or if we have some knowledge of the treatment mechanism, $\mathcal{M}$ is considered a semi-parametric model and we will incorporate such knowledge.

\section{Parameter of Interest and Identification}
First we define the potential outcome under the intervention to set the treatment to $a$ as \parencite{neyman1923} $Y_{a}=f_{Y}(U_{Y},a,W)$.  The blip function is then defined as $b_{P_{UX}}(W)  =  \mathbb{E}_{P_{UX}}[Y_{1}\vert W]-\mathbb{E}_{P_{UX}}[Y_{0}\vert W]$.  Our parameter of interest is a mapping from $\mathcal{M}^{F}$ to $R^{2}$ defined by $\Psi^{F}(P_{UX})=\mathbb{E}_{P_{UX}}\mathbb{I}(b_{P_{UX}}(W) \leq t)$ or the CDF of the blip. \\

We will impose the randomization assumption \parencite{Robins1986, Greenland1986}, $Y_{a}\perp A \vert W$ as well as positivity, $0 < E_P[A = a \mid W] < 1$ for all $a$ and $W$.  Defining $b_{P}(W)=\mathbb{E}_{P}[Y\vert A=1,W]-\mathbb{E}_{P}[Y\vert A=0,W]$ yields $b_{P_{UX}}(W)= b_{P}(W)$ and we can identify the parameter of interest as a mapping from the observed data model, $\mathcal{M}$, to $\mathbb{R}^{d}$ via the 

\[
\Psi(P)=\mathbb{E}_{P}\mathbb{I}(b(W) \leq t)\text{ for P\ensuremath{\in\mathcal{M}}}
\]

\smallskip{}

$\Psi$ is not pathwise differentiable \parencite{Vaart:2000aa} so instead we consider the
smoothed version of the parameter mapping, using kernel, $k$, with
bandwidth, $\delta$.  Here we will suppress $k$ in the notation for convenience:

{\footnotesize{}
\[
\Psi_{\delta,t}(P)=\mathbb{E}_{w}\int_{x}\frac{1}{\delta}k\left(\frac{x-t}{\delta}\right)\mathbb{I}(b(W) \leq x)dx=\int_{x}\frac{1}{\delta}k\left(\frac{x-t}{\delta}\right)F(x)dx
\]
}\bigskip{}
\bigskip{}

\textbf{NOTE}: We assume throughout this paper, $Pr(b(W)=x)=0$ for all values, $x$. In
other words, our blip distribution function is continuous. 

\section{Derivation of the Efficient Influence Curve of $\Psi_{\delta, t}(P)$}

\subsubsection{Tangent Space for Nonparametric Model}
The true data generating distribution, $P$, has density, $p$, which can be factorized as $p(o)=p_{Y}(y\vert a,w)g(a\vert w)p_{W}(w)$.
We consider the one dimensional set of submodels that pass through $P$ at $\epsilon=0$ \parencite{Vaart:2000aa} $\{P_\epsilon \text{ s.t. } p_\epsilon=(1+\epsilon S)p\vert \int SdP=0,\int S^2dP<\infty \}$.
The tangent space is the closure in $L^2$ norm of the set of scores, $S$, or directions for the paths defined above. We write:
\begin{eqnarray*}
T & = & \overline{\left\{ s(o)\vert\mathbb{E}S=0,\mathbb{E}S^2<\infty\right\}} \\
 & = & \overline{\{S(y\vert a,w)\vert\mathbb{E}_{P_{Y}}S=0,\mathbb{E}S^2<\infty\}}\oplus\overline{\{s(a\vert w)\vert\mathbb{E}_{P_{A}}S=0,\mathbb{E}S^2<\infty\}}\oplus\overline{\{S(w)\vert\mathbb{E}_{P_{W}}S=0,\mathbb{E}S^2<\infty\}}\\
  & = & T_{Y}\oplus T_{A}\oplus T_{W}
\end{eqnarray*}
For a non-parametric model, $T = L^{2}_{0}(P)$ forms a Hilbert space with inner product defined as $\langle f, g\rangle=\mathbb{E}_Pfg$.  Our notion of orthogonality now is $f \perp g$ if and only if $\langle f, g \rangle = 0$ and, therefore, the above direct sum is valid.  In other words, every score, $S$, can be written as $ \frac{d}{d\epsilon}log(p_{\epsilon})\vert_{\epsilon=0}=S(w,a,y)=S_Y(y\vert a,w)+S_A(a\vert w) + S_W(w)$ where, due to the fact $p_\epsilon=(1+\epsilon S)p=p_{Y\epsilon}p_{A\epsilon}p_{W\epsilon}$, it is easy to see $\frac{d}{d\epsilon}log(p_{Y\epsilon})\vert_{\epsilon=0}=S_Y(y\vert a,w)$, $\frac{d}{d\epsilon}log(p_{A\epsilon})\vert_{\epsilon=0}=S_A(a\vert w)$ and $\frac{d}{d\epsilon}log(p_{W\epsilon})\vert_{\epsilon=0}=S_W(w)$.  Furthermore we know that a projection of $S$ on $T_Y$ is  

\begin{eqnarray*}
S_Y(y \mid w, a) &= &S(w,a,y) - E[S(W, A, Y) \mid W = w, A = a] \\
& = & S(w,a,y) - \int S(w,a,y) p_Y(y \mid w, a) d\nu(y)\\
& = & \frac{d}{d\epsilon}log(p_{Y\epsilon})\vert_{\epsilon=0}
\end{eqnarray*}

\subsubsection{Efficiency Theory in brief}
The efficient influence curve at a distribution, $P$, for the parameter mapping, $\Psi_{\delta,t}$, is a function of the observed data, $O\sim P$, notated as $D^\star_{\Psi_{\delta,t}}(P)(O)$.  Its variance gives the generalized Cramer-Rao lower bound for the variance of any regular asymptotically linear estimator of $\Psi_{\delta,t}$ \parencite{Vaart:2000aa}.  For convenience we define the outcome model $\bar{Q}(A,W) = E_{P}[Y \mid A, W]$ and the treatment mechanism as $g(A \mid W) = E_{P}[A \mid W]$.  We will simplify the notation for the blip here as well, leaving off the subscript for the distribution so that $b(W) = E_{P}[Y \mid 1, W]-E_{P}[Y \mid 0, W]$.  As in van der Vaart, 2000, we define the pathwise derivative at $P$ along score, $S$, as 

\begin{equation}
\underset{\epsilon\rightarrow 0}{lim}\left(\frac{\Psi_{\delta,t}(P_\epsilon)-\Psi_{\delta,t}(P)}{\epsilon}\right)\longrightarrow \dot{\Psi}_{\delta,t}(S)
\end{equation}
We note to the reader, we imply a direction, $S$, when we write $P_{e}$, which has density $p(1+\epsilon S)$, but generally leave it off the notation as understood.\\

By the riesz representation theorem \parencite{riesz} for Hilbert Spaces, assuming the mapping in (1) is a bounded and linear functional on $T$, it can be written in the form of an inner product $\langle D^*(P),g \rangle$ where $D^*$ is a unique element of $T$, which we call the canonical gradient or efficient influence curve. Thus, in the case of a nonparametric model, the only gradient is the canonical gradient.  It is notable that the efficient influence curve has a variance that is the lower bound for any regular asymptotically linear estimator \parencite{Vaart:2000aa}. Since the TMLE, under conditions as discussed in this paper, asymptotically achieves variance equal to that of the efficient influence curve, the estimator is asymptotically efficient.\\

As a note to the reader: Our parameter mapping does not depend on the treatment mechanism, $g$, and also $T_{A}\perp T_{Y}\oplus T_{W}$ which, means our efficient influence curve must therefore be in $T_{Y}\oplus T_{W}$ for the nonparametric model.  Therefore, our efficient influence curve will have two orthogonal components in $T_Y$ and $T_W$ respectively. We have no component in $T_A$, which is why we need not perform a TMLE update of the initial prediction, $g_n$, of $g_0(A\vert W)$. Such also teaches us that for the semi-parametric model, where the treatment mechanism is known, the efficient influence function will remain the same.\\

\begin{thm}
Assume $k$ is lipschitz and smooth on $\mathbb{R}$. The efficient
influence curve for the parameter, $\Psi_{t,\delta}$, is given by
\end{thm}
{\footnotesize{}
\begin{eqnarray*}
\mathbf{D_{\Psi_{\delta, t}}^{\star}(P)(O)} & \mathbf{=} & \frac{-1}{\delta}\mathbf{k\left(\frac{b(W)-t}{\delta}\right)*\frac{2A-1}{g(A\vert W)}(Y-\bar{Q}(A,W))\mathbf{+\int\frac{1}{\delta}k\left(\frac{x-t}{\delta}\right)\mathbb{I}(b(W) \leq x)dx-\Psi_{\delta,t}}}
\end{eqnarray*}
}{\footnotesize \par}

PROOF: 

Define $\Phi(x) = 1/(1+exp(x))$.  We also  define $b_\epsilon = \mathbb{E}_{P_\epsilon} [Y \mid A = 1, W] - \mathbb{E}_{P_\epsilon}[Y \mid A = 1, W]$, where $P_\epsilon$ is defined via its density, $p_\epsilon = (1 + \epsilon S(o))p(o)$, and $p$ is the density of $P$.  $S$ is the so-called score function in Hilbert Space, $L^2_0(P)$, the completion (under the $L^2$ norm) of the space of mean 0 functions of finite variance.  We remind the reader that since our model is nonparametric, the tangent space is $L^2_0(P)$ \parencite{Vaart:2000aa}. We will now compute the pathwise derivative functional on $L^2_0(P)$, writing it as an inner product (covariance in the Hilbert Space $L^2(P)$), of the score, $S$, and the efficient influence curve, a unique element of the tangent space, $L^2_0(P)$.  We notate the efficient influence curve as indexed by the distribution, $P$, and as a function of the observed data, $O\sim P$: $D^*(P)(O)$.  By dominated convergence we have 
\[
\Psi_{\delta, t}(P)=\underset{h\rightarrow0}{lim}\mathbb{E}_{w}\int_{-1}^{1}\frac{1}{\delta}k\left(\frac{x-t}{\delta}\right)\Phi(\frac{b(W)-x}{h})dx
\]

{\footnotesize{}
\begin{eqnarray}
\underset{\epsilon\rightarrow0}{lim}\frac{\Psi_{\delta ,t}(P_{\epsilon})-\Psi_{\delta,t}(P)}{\epsilon} & = & \underset{\epsilon\rightarrow0}{lim}\frac{1}{\epsilon}\underset{h\rightarrow0}{lim}\mathbb{E}_{w}\int_{x}\frac{1}{\delta}k\left(\frac{x-t}{\delta}\right)\left(\Phi(\frac{b_{\epsilon}(W)-x}{h})-\Phi(\frac{b(W)-x}{h})\right)dx\\
 &  & +\mathbb{E}_{w}\left(\int_{x}\frac{1}{\delta}k\left(\frac{x-t}{\delta}\right)\mathbb{I}(b(W) \leq x)-\Psi_{t,\delta}(P)\right)S(O)dx\\
 &  & \text{let's ignore (2) for now}\\
 &  & \underset{\epsilon\rightarrow0}{lim}\frac{1}{\epsilon}\underset{h\rightarrow0}{lim}\mathbb{E}_{w}\int_{x}\frac{1}{\delta}k\left(\frac{x-t}{\delta}\right)\left(\frac{1}{h}\Phi^{\prime}(\frac{b(W)-x}{h})\left(b_{\epsilon}(W)-b(W)\right)\right)dx+\nonumber \\
 &  & +\underset{\epsilon\rightarrow0}{lim}\frac{1}{\epsilon}\underset{h\rightarrow0}{lim}\mathbb{E}_{w}\int_{x}\frac{1}{\delta}k\left(\frac{x-t}{\delta}\right)\left(\frac{1}{2h^{2}}\Phi^{(2)}\left(\zeta\left(\frac{x-b(W)}{h}\right)\right)\left(b_{\epsilon}(W)-b(W)\right)^{2}\right)dx\\
 & = & \underset{\epsilon\rightarrow0}{lim}\frac{1}{\epsilon}\underset{h\rightarrow0}{lim}\left(\mathbb{E}_{w}\int_{x}\frac{1}{\delta}k\left(\frac{x-t}{\delta}\right)\left(\frac{1}{h}\Phi^{\prime}(\frac{b(W)-x}{h})\left(b_{\epsilon}(W)-b(W)\right)\right)dx+R_{2,h,x}(b_{\epsilon},b)\right)
\end{eqnarray}
}{\footnotesize \par}

We can note that for $h(\epsilon)$ such that $\frac{\epsilon}{h^{2}(\epsilon)}\rightarrow0$
as $\epsilon\rightarrow0$, $\frac{R_{2}}{\epsilon}\rightarrow0$
because $R_{2}$ is order $\frac{\epsilon^{2}}{h^{2}}.$ To see this,
consider the convenient fact that $\Phi^{(2)}(x)$ is bounded. 

Let's now drop $\underset{\epsilon\rightarrow0}{lim}\frac{1}{\epsilon}$
for now and use integration by parts to compute a part of the integrand
in (5): {\scriptsize{}
\begin{eqnarray}
 &  & \mathbb{E}_{w}\underset{a\rightarrow\infty}{lim}\int_{t-a\delta}^{t+a\delta}\frac{1}{\delta}k\left(\frac{x-t}{\delta}\right)\frac{1}{h}\Phi^{\prime}(\frac{b(W)-x}{h})dx\left(b_{\epsilon}(W)-b(W)\right)\nonumber \\
 & = & \mathbb{E}_{w}\underset{a\rightarrow\infty}{lim}\left(\frac{-1}{\delta}k\left(\frac{x-t}{\delta}\right)\Phi(\frac{b(W)-x}{h})\biggr\vert_{t-a\delta}^{t+a\delta}+\int_{x}\frac{1}{\delta^{2}}k^{\prime}\left(\frac{x-t}{\delta}\right)\mathbb{E}_{w}\Phi(\frac{b(W)-x}{h})\right)\left(b_{\epsilon}(W)-b(W)\right)dx\nonumber \\
 & = & \mathbb{E}_{w}\underset{a\rightarrow\infty}{lim}\left(\frac{-1}{\delta}k\left(\frac{x-t}{\delta}\right)\Phi(\frac{b(W)-x}{h})\biggr\vert_{t-a\delta}^{t+a\delta}+\int\frac{1}{\delta^{2}}k^{\prime}\left(\frac{x-t}{\delta}\right)\mathbb{E}_{w}\left[\Phi(\frac{b(W)-x}{h})-\mathbb{I}(b(W) \leq x)\right]\right)\left(b_{\epsilon}(W)-b(W)\right)dx \nonumber\\
 &  & +\mathbb{E}_{w}\int_{x}\frac{1}{\delta^{2}}k^{\prime}\left(\frac{x-t}{\delta}\right)\mathbb{I}(b(W) \leq x)\left(b_{\epsilon}(W)-b(W)\right)dx\\
 &  & h\rightarrow0 \text{ and Dominated convergence} \implies \text{ 2nd term disappears.  k lipschitz} \implies \nonumber \\
 & = & \mathbb{E}_{w}\underset{a\rightarrow\infty}{lim}\frac{-1}{\delta}\mathbb{I}(b(W)\leq t+a\delta)k(a)+\frac{1}{\delta}\mathbb{I}(b(W) \leq  t-a\delta)k(-a))\\
 & &+ \frac{1}{\delta}\left( k(a) - k\left(\frac{max(b(W),t-a\delta)-t}{\delta}\right))\right)\mathbb{I}(b(W) \leq t-a\delta)\left(b_{\epsilon}(W)-b(W)\right)\\
 & = & \mathbb{E}_{w}\frac{-1}{\delta}k\left(\frac{b(W)-t}{\delta}\right)\left(b_{\epsilon}(W)-b(W)\right)\nonumber 
\end{eqnarray}
}{\scriptsize \par}

We can summarize as follows:

{\scriptsize{}
\begin{eqnarray*}
\underset{\epsilon\longrightarrow0}{lim}\frac{\Psi_{\delta,t}(P_{\epsilon})-\Psi_{\delta,t}(P)}{\epsilon} & = & \underset{\epsilon\rightarrow0}{lim}\frac{1}{\epsilon}\mathbb{E}_{w}\left(\frac{-1}{\delta}k\left(\frac{b(W)-t}{\delta}\right)\left(b_{\epsilon}(W)-b(W)\right)\right)+\\
 &  & +\underset{\epsilon\rightarrow0}{lim}\underset{h(\epsilon)\rightarrow0}{lim}\mathbb{E}_{w}\frac{R_{2,h,x}(b_{\epsilon},b)}{\epsilon}\\
 &  & +\mathbb{E}_{w}\left(\int_{x}\frac{1}{\delta}k\left(\frac{x-t}{\delta}\right)\mathbb{I}(b(W) \leq x)-\Psi_{t,\delta}(P)\right)S(O)dx
\end{eqnarray*}
}{\scriptsize \par}

As previously stated, the second term disappears by easy choice of
$h$. 

\begin{align*}
\frac{d}{d\epsilon}p_{Y\epsilon}(y\vert a=0,w)\vert_{\epsilon=0} & =p_{Y}(y\vert a,w)S_{Y}(y\vert a,w)\\
 & =p_{Y}(y\vert a,w)\left(S(o)-\mathbb{E}[S(o)\vert a,w]\right)
\end{align*}

We then compute the pathwise derivative along $S$ at $\epsilon=0$:
\begin{align*}
\underset{\epsilon\rightarrow0}{lim}\frac{1}{\epsilon}\mathbb{E}_{w}\left(\frac{-1}{\delta}k\left(\frac{b(W)-t}{\delta}\right)\left(b_{\epsilon}(W)-b(W)\right)\right) & =\\
\int\left(\frac{-1}{\delta}k\left(\frac{b(w)-t}{\delta}\right)\int\left(p_{\epsilon}(y\vert a=1,w)-\frac{d}{d\epsilon}p_{Y\epsilon}(y\vert a=0,w)\right)d\nu(y)\right)p_{W}(w)d\nu(w) & =\\
\int\left(\frac{-1}{\delta}k\left(\frac{b(w)-t}{\delta}\right)\int\int\frac{2a-1}{g(a\vert w)}yp_{Y\epsilon}(y\vert a,w)S_{Y}(o)d\nu(y)\right)g(a\vert w)p_{W}(w)d\nu(a,w) & =\\
\int\left(\frac{-1}{\delta}k\left(\frac{b(w)-t}{\delta}\right)\int\int\frac{2a-1}{g(a\vert w)}yp_{Y\epsilon}(y\vert a,w)\left(S(o)-\mathbb{E}[S\vert a,w]\right)d\nu(y)\right)g(a\vert w)p_{W}(w)d\nu(a,w) & =\\
\int\frac{-1}{\delta}k\left(\frac{b(w)-t}{\delta}\right)\frac{2a-1}{g(a\vert w)}S(o)p(o)d\nu(o)-\int\frac{1}{\delta}\bar{Q}(a,w)S(o)p(o)d\nu(o) & =\\
\biggr\langle\frac{-1}{\delta}k\left(\frac{b(w)-t}{\delta}\right)\frac{2A-1}{g(A\vert W)}(Y-\bar{Q}(A,W)),S(O)\biggr\rangle_{L_0^2(P)}
\end{align*}

Thus we finally get: 
\begin{align*}
\underset{\epsilon\longrightarrow0}{lim}\frac{\Psi_{\delta,t}(P_{\epsilon})-\Psi_{t,\delta}(P)}{\epsilon} & =\mathbb{E}_{w}\left(\int_{x}\frac{1}{\delta}k\left(\frac{x-t}{\delta}\right)\mathbb{I}(b(W) \leq x)-\Psi_{\delta,t}(P)\right)S(O)dx+\\
 & +\biggr\langle\frac{-1}{\delta}k\left(\frac{b(w)-t}{\delta}\right)\frac{2A-1}{g(A\vert W)}(Y-\bar{Q}(A,W)),S(O)\biggr\rangle_{L_0^2(P)}\\
 & =\langle D_{\Psi_{\delta,t}}^{\star}(P),S\rangle_{L_0^2(P)}
\end{align*}

where

\begin{eqnarray*}
\mathbf{D_{\Psi_{\delta, t}}^{\star}(P)(O)} & \mathbf{=} & \mathbf{\frac{-1}{\delta}k\left(\frac{b(W)-t}{\delta}\right)*\frac{2A-1}{g(A\vert W)}(Y-\bar{Q}(A,W))\mathbf{+\int\frac{1}{\delta}k\left(\frac{x-t}{\delta}\right)\mathbb{I}(b(W) \leq x)dx-\Psi_{\delta, t}}}
\end{eqnarray*}

And this is the efficient influence curve since the canonical gradient
is the only gradient for a non-parametric model where the closure
of the set of scores is all of $L_0^2(P)$.  \\

\hspace{5.8in}QED

\section{The Targeted Maximum Likelihood Estimator, TMLE}
We will employ the notation, $P_{n}f(O)$ to be the empirical average of function, $f(\cdot)$, and $Pf(O)$ to be $\mathbb{E}_{P}f(O)$.  Define a loss function, $L(P)(O)$, which is a function of the observed data, O, and indexed at the distribution on which it is defined, $P$, such that $E_{P_0} L(P)(O)$ is minimized at the true observed data distribution, $P=P_0$. The targeted maximum likelihood (TML) estimating procedure maps an initial estimate, $P_{n}^{0}\in \mathcal{M}$, of the true data generating distribution to $P_{n}^{\star}\in \mathcal{M}$ such that $P_{n}L(P_{n}^{\star})\leq P_{n}L(P_{n}^{0})$ and such that $P_{n}D^{\star}_{\Psi_{\delta, t}}(P_{n}^{\star})=0_{d\times1}$, where $d$, in this case, is the number of points on the CDF.  $P_{n}^{\star}$ is called the TMLE of the initial estimate $P_{n}^{0}$ \parencite{Laan:2006aa, Laan:2011aa}.  For convenience, we define, $\bar{Q}_0(A,W) = E_{P_0}[Y \mid A, W]$ and its initial estimate, $\bar{Q}_n^0(A,W)$ and we will use $g_0(A \mid W) = E_{P_0}[A \mid W]$ with corresponding initial estimate, $g_n$.  The initial estimate of the distribution of $W$ is denoted $Q_{W,n}$, the empirical distribution of $W$, with density, $q_{W,n}$.   For this paper, the TMLE procedure only adjusts the initial estimate of the outcome regression, leaving $g_n$ and $q_{W,n}$ as is.  Thus we will only update $\bar{Q}_n^0(A,W)$ to its TMLE, $\bar{Q}_n^*(A,W)$.  \\

To perform the TMLE updating procedure, we may find an element of either a universal least favorable submodel (ulfm), a least favorable submodel (lfm), both defined in van der Laan and Gruber, 2016, or a canonical least favorable submodel \parencite{clfm}.  Both clfm and ulfm use a single dimensional submodel where as the lfm is of dimension, $d$, and identical to a clfm if $d=1$.  The ulfm has the advantage of not relying on iteration as explained in van der Laan and Gruber, 2016, but here we did not notice an appreciable difference in performance so we used the faster clfm procedure.  To construct a clfm, ulfm or lfm, one needs to know the efficient influence curve, which is given by 

{\footnotesize{}
\begin{eqnarray*}
\mathbf{D_{\Psi_{\delta, t}}^{\star}(P)(O)} & \mathbf{=} & \frac{-1}{\delta}\mathbf{k\left(\frac{b(W)-t}{\delta}\right)*\frac{2A-1}{g(A\vert W)}(Y-\bar{Q}(A,W))\mathbf{+\int\frac{1}{\delta}k\left(\frac{x-t}{\delta}\right)\mathbb{I}(b(W) \leq x)dx-\Psi_{\delta,t}}}
\end{eqnarray*}
}{\footnotesize \par}
where we estimate the CDF of blip at a given blip value, $t$, using kernel, $k$, and bandwidth $\delta$ \parencite{blipCDFtech}.  The CV-TMLE algorithm by the author \parencite{cvtmle} simplifies the originally formulated CV-TMLE algorithm by Zheng and van der Laan, 2010 and, in this case, turns out to be the same estimator if we use a pooled regression to fit the fluctuation parameter.  The TMLE updating procedure is implemented in the software packages blipCDF \parencite{blipCDF} and \parencite{tmle3}.  Here we will provide for readers more familiar with TMLE, only the so-called clever covariate \parencite{Laan:2006aa} for $\Psi_{h,t}$, but the reader may consult Levy, 2018c for a detailed algorithm.  

\[
H(A,W) = \frac{-1}{\delta}k\left(\frac{b(w)-t}{\delta}\right)\frac{2A-1}{g(A\vert W)}
\]  

If we are simultaneously estimating $d$ points, $t_1, t_2, ..., t_d$ on the CDF curve,  we will have a $d-dimensional$ clever covariate:

\[
(H_1(A,W), H_2(A,W),...,H_d(A,W)) = \frac{1-2A}{\delta g(A\vert W)}\left(k\left(\frac{b(w)-t_1}{\delta}\right), k\left(\frac{b(w)-t_2}{\delta}\right),...,k\left(\frac{b(w)-t_d}{\delta}\right)\right)
\]  

The TMLE procedure yields $\bar{Q}_n^*(A,W)$ and our estimator is then a plug-in, using the empirical distribution, $Q_{W,n}$:

\[
{\Psi}_{\delta,t}(P_n^*) = \frac{1}{n}\sum_{i=1}^n \frac{1}{\delta} \int k\left(\frac{x - t}{\delta} \right) \mathbb{I}(b_n^*(W_i) \leq x) dx
\]

where $b_n^*(W_i) = \bar{Q}_n^*(1,W) - \bar{Q}_n^*(0,W)$, the blip function estimate.  For simultaneously estimating many points on the CDF of the blip, the TMLE procedure yields a common outcome model for all $t$-values, $t_1, t_2, ..., t_d$, which has the advantage of preserving monotonicity.   

\subsubsection{Software}
The TMLE is implemented in the software packages blipCDF \parencite{blipCDF} and \parencite{tmle3}. 

\section{TMLE conditions}
By solving the efficient influence curve equation with our TMLE update, we can then write a second order expansion, $\Psi_{\delta}(P_{n}^{\star})-\Psi_{\delta}(P_{0})=(P_{n}-P_{0})D^{\star}_{\Psi_{\delta}}(P_{n}^{\star})+R_{2}(P_{n}^{\star},P_{0})$. We then arrive at the following three conditions (for fixed bandwidth, $\delta$) that guarantee asymptotic efficiency for this estimator \parencite{Laan:2006aa, Laan:2011aa}, the first of which is not required for CV-TMLE \parencite{Zheng:2010aa}. Thus CV-TMLE is our preferred estimator, since it requires less conditions on our machine learning, enabling a more aggressive approach to fitting the treatment mechanism and the outcome model.  

\subsection{TMLE Conditions and Asymptotic Efficiency}
 We refer the reader to Targeted Learning Appendix \parencite{Laan:2011aa} as well as \parencite{Laan:2015aa,Laan:2015ab, Laan:2006aa} for a more detailed look at the theory of TMLE.  For convenience, we will summarize some of the main results for the reader.  

\subsubsection{Conditions for Asymptotic Efficiency}
Define the norm $\Vert f \Vert_{L^{2}(P)} = \sqrt{\mathbb{E}_{P}f^{2}}$. Assume the following TMLE conditions:

\begin{enumerate}
\item
$D^{\star}_{\Psi_{\delta,t}}(P_{n}^{\star})$ is in a P-Donsker class. This condition can be dropped in the case of using CV-TMLE \parencite{Zheng:2010aa}. We show the advantages to CV-TMLE in our simulations.  

\item
$R_{2}(P_n^*,P_0)$ is $o_{p}(1/\sqrt{n})$ for all $j$.
\item
$D^{\star}_{\Psi_{\delta,t}}(P_{n}^{\star})\overset{L^{2}(P_{0})}{\longrightarrow} D^{\star}_{\Psi_{\delta, t}}(P_{0})$ 

\end{enumerate}

then $\sqrt n(\Psi_{\delta,t}(P_{n}^{\star})-\Psi_{\delta,t}(P_{0})) \overset{D}{\implies} N[0, var_{P_0}(D^{\star}_{\Psi}(P_{0})]$.  Our plug-in TMLE's and CI's are given by 

$$\Psi_{\Psi_{\delta,t}}(P_{n}^{\star})\pm z_{\alpha}*\frac{\widehat{\sigma}_n(D_{\Psi_{\delta,t}}^{\star}(P_{n}^{\star}))}{\sqrt{n}}$$ 

Under the above conditions, these confidence bands will be as small as possible for any regular asymptotically linear estimator at significance level, $1-\alpha$, where $Pr(\vert Z \vert \leq z_{\alpha})=\alpha$ for Z standard normal and $\widehat{\sigma}_n(D_{j}^{\star}(P_{n}^{\star}))$ is the sample standard deviation of $\{D_{j}^{\star}(P_{n}^{\star})(O_i) \mid i \in 1:n \}$ \parencite{Laan:2006aa}.  Note, that if the TMLE conditions hold for the initial estimate, $P_n^0$, then they will also hold for the updated model, $P_n^{\star}$ \parencite{Laan:2015aa}, thereby placing importance on our ensemble machine learning in constructing $P_n^0$.  For simultaneous confidence intervals, we refer the reader to Levy, van der Laan et al., 2018, for the method which leverages the efficient influence curve approximation to form confidence bounds that simultaneously cover the parameter values at a given significance level.  Such inference is as tight as possible and certainly tighter than a standard bonferroni correction \parencite{bonferroni}. \\  

\section{The Remainder Term for a TMLE Plug-in Estimator of $\Psi_{\delta,t}(P_{0})$}

In this section we will prove the remainder term of the previous section is $\frac{1}{\delta}O\left(\Vert g-g_{0}\Vert_{L_{P_{0}}^{2}}\Vert\bar{Q}-\bar{Q}{}_{0}\Vert_{L_{P_{0}}^{2}}\right)+\frac{1}{\delta}O\left(\Vert b-b_{0}\Vert_{\infty}^{2}\right)$, assuming WLOG that the support of the kernel is $[-1,1]$.   

\begin{lem}
Assume lipschitz $F_{0}=1-S$, where $S(t)=\mathbb{E}\mathbb{I}(b_{0}(W)>t)$
and assume WLOG the support of the kernel is $[-1,1]$ 

then $P_{0}\mathbb{I}(b_{0}(W)>t+\delta,b(W)<t+\delta)=O(\Vert b_{0}-b\Vert_{\infty}$ 

proof: 
\end{lem}
{\footnotesize{}
\begin{eqnarray*}
P_{0}\mathbb{I}(b_{0}(W)>t+\delta,b(W)<t+\delta) & = & P_{0}\mathbb{I}(b_{0}(W)>t+\delta,b(W)<t+\delta)\mathbb{I}(b_{0}(W)-b(W)>b_{0}(W)-(t+\delta))\\
 & \leq & P_{0}\mathbb{I}(b_{0}(W)>t+\delta,b(W)<t+\delta)\mathbb{I}(\Vert b_{0}-b\Vert_{\infty}>b_{0}(W)-(t+\delta))\\
 & \leq & Pr(t+\delta<b_{0}(W)<\Vert b_{0}-b\Vert_{\infty}+t+\delta)\\
\text{Lipschitz \ensuremath{\implies}} & \leq & L\Vert b_{0}-b\Vert_{\infty}+O(\Vert b_{0}-b\Vert_{\infty}^{2})
\end{eqnarray*}
 }{\footnotesize \par}

\hspace{5.8in}\textbf{QED}
\begin{thm}
The remainder term, $R_{2}(P_{\epsilon},P)=P_{0}D^{*}(P)+\Psi(P)-\Psi(P_{0})$,
is $O\left(\Vert b_{0}-b\Vert_{\infty}^{2}\right)$ 

Proof: 
\end{thm}
\textbf{\scriptsize{}
\begin{eqnarray}
R_{2}(P_{0}P) & = & P_{0}D^{*}(P)+\Psi(P)-\Psi(P_{0})\nonumber \\
 & = & P_{0}\left[\frac{-1}{\delta}k\left(\frac{b(W)-t}{\delta}\right)\frac{2A-1}{g(A\vert W)}\left(Y-\bar{Q}(A,W)\right)+\int\frac{1}{\delta}k\left(\frac{x-t}{\delta}\right)\mathbb{I}\left(b(W)>x\right)dx-\int_{x}\frac{1}{\delta}k\left(\frac{x-t}{\delta}\right)\mathbb{I}(b_{0}(W)>x)dx\right]\nonumber \\
 & = & P_{0}\frac{-1}{\delta}k\left(\frac{b(W)-t}{\delta}\right)\left(\frac{2A-1}{g(A\vert W)}\left(Y-\bar{Q}(A,W)\right)\right)+P_{0}\int\frac{1}{\delta}k\left(\frac{x-t}{\delta}\right)\left(\mathbb{I}\left(b(W)>x\right)-\mathbb{I}(b_{0}(W)>x)\right)dx\\
 & = & \frac{-1}{\delta}P_{0}\left[k\left(\frac{b(W)-t}{\delta}\right)\left(\left(\frac{g_{0}(1\vert W)}{g(1\vert W)}\right)\left(\bar{Q}_{0}(1,W)-\bar{Q}(1,W)\right)-\left(\frac{g_{0}(0\vert W)}{g(0\vert W)}\right)\left(\bar{Q}_{0}(0,W)-\bar{Q}(0,W)\right)\right)\right]+\\
 &  & \frac{1}{\delta}P_{0}\int_{b(W)}^{b_0(W)}k\left(\frac{x-t}{\delta}\right)dx\\
 & = & \frac{-1}{\delta}P_{0}\left[k\left(\frac{b(W)-t}{\delta}\right)\left(\left(\frac{g_{0}(1\vert W)}{g(1\vert W)}-1\right)\left(\bar{Q}_{0}(1,W)-\bar{Q}(1,W)\right)-\left(\frac{g_{0}(0\vert W)}{g(0\vert W)}-1\right)\left(\bar{Q}_{0}(0,W)-\bar{Q}(0,W)\right)\right)\right]\\
 &  & +\frac{1}{\delta}P_{0}\left[\int_{b(W)}^{b_0(W)}k\left(\frac{x-t}{\delta}\right)dx+k\left(\frac{b(W)-t}{\delta}\right)\left(b(W)-b_0(W)\right)\right]
\end{eqnarray}
}{\scriptsize \par}

Clearly (12) will disappear if $g_{0}$ is known. Otherwise the term
is $\frac{1}{\delta}\Vert g-g_{0}\Vert_{L_{p_{0}}^{2}}\Vert\bar{Q}-\bar{Q}_{0}\Vert_{L_{P_{0}}^{2}}$
by cauchy-schwarz. 

Now let's take a look at (13):
\textbf{\scriptsize{}
\begin{eqnarray}
 &  & \frac{1}{\delta}P_{0}\left[\int_{b_{0}(W)}^{b(W)}k\left(\frac{x-t}{\delta}\right)dx+k\left(\frac{b(W)-t}{\delta}\right)\left(b_{0}(W)-b(W)\right)\right]
\end{eqnarray}
}{\scriptsize \par}

We can divide the $W$ space into disjoint parts and integrate: 

a) $t-\delta<b_{0}(W)<t+\delta$:

Assuming $F_{0}$ is lipschitz, we have as follows:

{\footnotesize{}
\begin{eqnarray}
 &  & \frac{1}{\delta}P_{0}\mathbb{I}(t-\delta<b_{0}(W)\leq t+\delta)*\left[\int_{b_{0}(W)}^{b(W)}k\left(\frac{x-t}{\delta}\right)dx+k\left(\frac{b(W)-t}{\delta}\right)\left(b_{0}(W)-b(W)\right)\right]\nonumber \\
 &  & \text{taylor expanding }\mbox{k\ensuremath{\left(\frac{x-t}{\delta}\right)} about }\frac{b(W)-t}{\delta}\text{ we obtain}\\
 &  & \frac{1}{\delta}P_{0}\mathbb{I}(t-\delta<b_{0}(W)\leq t+\delta)*\int_{b_{0}(W)}^{b(W)}k^{\prime}\left(\gamma\left(x,b(W),\delta\right)\right)\left(\frac{x-b(W)}{\delta}\right)\nonumber \\
 &  & \text{where }\mbox{\ensuremath{\gamma\left(x,b(W),\delta\right)} is an intermediary point}\\
 & \leq & P_{0}\mathbb{I}(t-\delta<b_{0}(W)\leq t+\delta)*\left(\frac{b_{0}(W)-b(W)}{\delta}\right)^{2} = \frac{1}{\delta}\int \left( (b_0 - b)^2 \mathbb{I}(t-\delta < b_0 < t + \delta)dP_{b_0,b}(b_0,b)\right)\nonumber\\
\text{or} & \leq & \frac{1}{\delta^{2}}\left(F_{0}(t+\delta)-F_{0}(t-\delta)\right)*\Vert b_{0}(W)-b(W)\Vert_{\infty}^{2}\nonumber \\
 &  & \text{employing the Lipschitz condition for \ensuremath{F_{0}} we arrive at }\nonumber \\
 & \leq & \frac{1}{\delta}O(\Vert b-b_{0}\Vert_{\infty}^{2})
\end{eqnarray}
}{\footnotesize \par}

b) $b_{0}(W)>t+\delta,b(W)\leq t+\delta$

{\footnotesize{}
\begin{eqnarray}
 &  & \frac{1}{\delta}P_{0}\mathbb{I}(b_{0}(W)>t+\delta,b(W)\leq t+\delta)*\left[\int_{b_{0}(W)}^{b(W)}k\left(\frac{x-t}{\delta}\right)dx+k\left(\frac{b(W)-t}{\delta}\right)\left(b_{0}(W)-b(W)\right)\right]\\
 & = & \frac{1}{\delta}P_{0}\left[\int_{b_{0}(W)}^{b(W)}\left(k\left(\frac{b(W)-t}{h}\right)+\left(\frac{x-t}{h}\right)k^{\prime}\left(\gamma\left(b(W),x,t,h\right)\right)\right)dx+k\left(\frac{b(W)-t}{h}\right)\left(b_{0}(W)-b(W)\right)\right]*\nonumber \\
 &  & \mathbb{I}(b_{0}(W)>t+\delta,b(W)\leq t+\delta)\nonumber \\
\text{} & \leq & \frac{1}{\delta}P_{0}\frac{\mathbb{I}(t+\delta<b_{0}(W)<t+\delta+\Vert b_{0}-b\Vert_{\infty})}{\delta}*C(b_{0}(W)-b(W))^{2}\nonumber \\
 & = & \frac{C}{\delta}\int\left(b_{0}-b\right)^{2}\frac{\mathbb{I}(t+\delta<b_{0}(W)<t+\delta+\Vert b_{0}-b\Vert_{\infty})}{\delta}dP_{b,b_{0}}(b,b_{0})\nonumber \\
 & or & \frac{1}{\delta}O\Vert b_{0}-b\Vert_{\infty}^{2}\text{ from (18) if we employ lemma 4.1}\nonumber 
\end{eqnarray}
}{\footnotesize \par}

c) $b_{0}(W)\leq t-\delta,b(W)>t-\delta$ This region follows identically
to b). 

d) for the cases, $b(W)$ and $b_{0}(W)<t-\delta$ or $b(W)$ and
$b_{0}(W)>t+\delta$, we can notice 

$\left[\int_{b_{0}(W)}^{b(W)}k\left(\frac{x-t}{\delta}\right)dx+k\left(\frac{b(W)-t}{\delta}\right)\left(b_{0}(W)-b(W)\right)\right]=0$.

\subsubsection{Variance is of order $1/\delta$: }
\begin{thm}
The asymptotic variance of our TMLE estimator is of order $1/\delta$ if we satisfy the TMLE conditions of section 6.
\end{thm}
We will compute the variance of the efficient influence curve and show it is of order $1/\delta$, thus proving the point.    

\begin{align*}
\mathbb{E}\frac{1}{\delta^{2}}k^{2}\left(\frac{b(W)-t}{\delta}\right)*\left[\frac{2A-1}{g(A\vert W)}(Y-\bar{Q}(A,W))\right]^{2}\leq & \mathbb{E}\mathbf{\frac{C}{\delta^{2}}k^{2}\left(\frac{b(W)-t}{\delta}\right)}\\
= & \mathbb{E}\frac{C}{\delta^{2}}\frac{d}{db(W)}\int_{-\infty}^{\infty}\mathbb{I}(b(W) \leq x)k^{2}\left(\frac{x-t}{\delta}\right)dx\\
= & \mathbb{E}\frac{C}{\delta}\frac{d}{db(W)}\int_{-\infty}^{\infty}\mathbb{I}\left(\frac{b(W)-t}{\delta} \leq y\right)k^{2}\left(y\right)dy\\
= & O\left(1/\delta\right)
\end{align*}

Now, we assume $k$ has finite support, WLOG, between $[-1,1]$ and
that $\vert k^{2}(x)\vert\leq M$

\begin{align*}
\mathbb{E}\mathbf{\frac{C}{\delta^{2}}k^{2}\left(\frac{b(W)-t}{\delta}\right)}\leq & \mathbb{E}\mathbf{\frac{C_{1}}{\delta^{2}}\mathbb{I}\left(-1\leq\frac{b(W)-t}{\delta}\leq1\right)}\\
= & \mathbb{E}\mathbf{\frac{C_{1}}{\delta^{2}}\mathbb{I}\left(-\delta+t\leq b(W)\leq\delta+t\right)}\\
= & \frac{C_{1}}{\delta^{2}}\left[F\left(\delta+t\right)-F\left(-\delta+t\right)\right]\\
\overset{Lipschitz}{\leq} & \frac{C_{2}}{\delta}
\end{align*}

\hspace{6in} QED

\subsubsection{Order of the bias}

\begin{align}
\mathbb{E}_{P_{W}}I(b(W) \leq t)-\mathbb{E}_{P_{W}}\int\frac{1}{\delta}k\left(\frac{x-t}{\delta}\right)I(b(W) \leq x)dx & \overset{fubini}{=}\nonumber \\
F(t)-\int\frac{1}{\delta}k\left(\frac{x-t}{\delta}\right)F(x)dx & =\nonumber \\
F(t)-\int k\left(y\right)F(y\delta+t)dy & =\nonumber \\
\int k\left(y\right)\left[F(t)-F(y\delta+t)\right]dy & =\\
\int k\left(y\right)\left[\sum_{i=1}^{\infty}F^{(i)}(t)(y\delta)^{i}/i!\right]dy
\end{align}

where (20) follows, assuming smoothness of the blip CDF function. This
will make the order of the bias dependent on the order of the kernel,
$k$. Without smoothness, a lipschitz condition on the blip CDF assures
that the bias is of order $\delta$ from (19). 

\subsubsection{Generating Kernels}
We generate a kernel of order $K+1$ as follows.  by generating symmetric polynomial kernels of finite support, the integration can be obtained via an explicit formula and is thus much faster and more accurate than numerical integration. We form polynomials of the form $k(x) = \sum_{i=0}^{K+2}a_i x^{2i}$ where the support of the kernel is from $-R$ to $R$.  The kernel $k(\cdot)$ is of course orthogonal to any odd power.  To make it order $K+1$ for $K$ an even positive number, we solve the following equations.    
\begin{enumerate}
\item
make sure the kernel is 0 at the end pts of the support: $$\sum_{i=0}^{K+2}a_i R^{2i}=0$$
\item
make sure the kernel has derivative 0 at the end pts of the support in consideration of the remainder term analysis:
$$\sum_{i=0}^{K+1}2a_i R^{2i+1}=0$$
\item
To enforce the necessary orthogonality, we solve for $K>0$ and each $r$ in the $2,4,...,2K$ $$2\sum_{i=0}^{K+2}a_i \frac{R^{2i+1+r}}{2i+1+r}=0$$
\end{enumerate}

\subsubsection{The Remainder Term Condition for Fixed Bandwidth}
Considering section 5 remainder term results, if $Vert\bar{Q}_{n}^{*}-\bar{Q}_{0}\Vert_{L^{2}(P_{0})}=o_P(n^{r_{\bar{Q}}})$ and the second is $\Vert g_{n}-g_{0}\Vert_{L^{2}(P_{0})}=o_P(n^{r_g})$, then $r_{\bar{Q}} + r_g \leq -0.5$ will partially satisfy the TMLE remainder term condition 2 of section 6.  However, we also need $\frac{C}{\delta}\Vert b-b_{0}\Vert_{\infty} = o_P(n^{-0.25})$, in order to guarantee the CV-TMLE estimator is unbiased and asymptotically efficient.  Hence, CDF of the blip estimation is not doubly robust and the $L^\infty$ norm sufficient condition is more demanding than if we have the same requirement  but with the $L^2(P_0)$ norm, as for the variance of the treatment effect (VTE) \parencite{blipvar}.  The highly adaptive lasso \parencite{Laan:2015ab} guarantees the latter but not the former.  However, $C \Vert b_{0}-b\Vert_{\infty}^{2}$ forms an upper bound for the remainder term where as $-\Vert b_{0}-b\Vert_{L^2(P_0)}^{2}$ is the exact remainder term for VTE, so there might be cases where the remainder term requirement for the CDF of the blip is easier to fulfill.

\subsubsection{Allowing the Bandwidth to Vanish and Conditions for Asymptotic Normality}
The order of the variance is $1/\delta$ and the order of the bias is $\delta^J$, where $J$ is the degree of the first non-zero moment of the kernel \parencite{blipCDFtech} or so-called order of the kernel.  Thus our situation of kernel and bandwidth selection resembles the case for a standard kernel density estimator.  However, we are additionally burdened with a second order remainder term that contains $1/\delta$ as a factor so we need to establish conditions under which this term does not blow up with increasing sample size.  If we allow the bandwidth $\delta_n$ to go to 0 so as to estimate the parameter, $\Psi(P)=\mathbb{E}_{}\mathbb{I}(b(W)>t)\text{ for P\ensuremath{\in\mathcal{M}}}$, then our remainder term teaches us that even if we know the true treatment mechanism, as in the case of a randomized trial, we require the sufficient condition 

\begin{equation}
\frac{C}{\sqrt{\delta_n}}\Vert \bar{Q}_0 - \bar{Q}_n^0 \Vert_{\infty} = o_P(n^{-0.25})
\end{equation} 

where we consider $\delta_n$ going to 0 as $n$ goes to infinity.  Considering the order of the bias and variance, we see that to minimize the MSE, our optimal bandwidth is $O(n^{\frac{-1}{2J+1}})$, where $J$ is the degree of the 1st non-zero moment or so-called order of the kernel.  The remainder condition would then require 

\begin{equation}
\Vert \bar{Q}_0 - \bar{Q}_n^0 \Vert_\infty = o_P\left( n^{-\frac{2J+3}{4(2J+1)}} \right)
\end{equation}

Thus, perhaps higher order kernels can be useful in that they require a rate closer to $n^{-0.25}$ for the sup norm bias.  The highly adaptive lasso or so-called HAL \parencite{Laan:2015aa}, guarantees $\Vert \bar{Q}_0 - \bar{Q}_n^0 \Vert_{L^2(P_0)} = O_P (n^{-1/4-1/8(d+1)})$, where $d$ is the dimension of $(A,W)$, but the reader may notice this rate is for the less stringent $L^2$ norm.  Even if we were to have a remainder condition in terms of the $L^2$ norm rather than the $L_\infty$ norm, a high dimensional $(A,W)$ would necessitate a kernel of order greater than $\frac{4d+3}{2}$.  

\section{Simulations}
\subsection{Well-specified Models}
For well-specified logistic models where the data generating system is given by the following: $W$ is a random normal, $Pr(A=1 \mid W) = g(A \mid W) = expit(.2+.2*W)$ and $E[Y \mid A,W] = expit(A + 2.5*A*W + W)$.  The TMLE's using the MLE as an initial estimate performed very well, with normal sampling distributions, nominal coverage (93\% or higher), as expected, and did so for all kernels if we used bandwidth $n^{-1/(2J+1)}$ where $J$ is the order of the kernel we let $n$ attain values of 1000, 2500, 5000, 10000, 25000 and 50000.  The MSE was lowest for the well-specified MLE plug-in, also as expected, but not appreciably.  Hence TMLE is a fine estimator in the near impossible situation where we might get a parametric form correct. 



\subsubsection{A Method for Choosing Bandwidth for a Given Kernel}
We would like to form confidence bounds for the non-pathwise differentiable parameter, $\Psi(P)=\mathbb{E}_{P}\mathbb{I}(b(W) \leq t)\text{ for P\ensuremath{\in\mathcal{M}}}$, and propose using some of the concepts in Chapter 25 of Targeted Learning in Data Science: Causal Inference for Complex Longitudinal Studies\parencite{TLII}.  We start with a largest bandwidth of size $n^{1/2J+1}$ where $J$ is the order of the kernel.  Then we divide the bandwidth into 20 equal increments from $n^{1/2J+1}/20, 2n^{1/2J+1}/20, ..., n^{1/2J+1} $.  We then find the smallest set of 5 or more consecutive bandwidths that are monotonic estimates with respect to the bandwidth.  If no such 5 or more consecutive bandwidths are found then we choose the bandwidth $n^{1/2J+1}$.  Let us call the consecutive bandwidth sequence, $B_c = \{h_1, ..., h_c\}$, where $h_1$ is the smallest.  We also monotonize the variance so as to force it to be increasing as the bandwidth gets smaller.  We then form confidence intervals using the monotonized variance for each bandwidth in $B_c$.  If the sequence of estimates is decreasing (increasing) as bandwidth decreases (for bandwidths in $B_c$), then we choose the confidence interval with the minimum (maximum) right (left) bound.  The idea is that we are minimizing the MSE via this choice, assuming that our region, $B_c$ represents the monotonicity as the bandwidth approaches 0.  We still need to refine the theory as our increments for the bandwidth (20 in this case) and definition of being monotonic (5 consecutive or more as described above) are somewhat arbitrary.  On the positive side, we noticed coverage of the smoothed parameter maintained nominal levels as $n$ grew to 50,000 when applying our bandwidth selector.  Figure 1 below displays the heuristic behind our bandwidth selector.  

\begin{table}[!htbp] \centering 
  \caption{coverage of smoothed parameter, kernel is order 10} 
  \label{} 
  \begin{scriptsize}
\begin{tabular}{@{\extracolsep{5pt}} cccccc}
\hspace{2cm} n = 1000  \hspace{.25cm} &  \hspace{.3cm} n = 2500 \hspace{.25cm} &  \hspace{.3cm} n = 5000 \hspace{.25cm} & \hspace{.3cm} n = 10000\hspace{.25cm}  
& \hspace{.3cm} n = 25000 \hspace{.25cm}  &  \hspace{1cm} n = 50000 \\ 
\end{tabular}
\begin{tabular}{@{\extracolsep{5pt}} ccccccccccccc} 
blip & meth & fixed & meth &   fixed & meth  &  fixed  & meth &  fixed
& meth &  fixed & meth & fixed \\ 
\hline \\[-1.8ex]       
$ -0.145 $ & $0.907$ & $0.947$ & $0.920$ & $0.949$ & $0.916$ & $0.941$ & $0.935$ & $0.948$ & $0.944$ & $0.949$ & $0.944$ & $0.948$ \\ 
$-0.085 $ &$0.911$ & $0.953$ & $0.950$ & $0.946$ & $0.939$ & $0.934$ & $0.958$ & $0.962$ & $0.942$ & $0.947$ & $0.955$ & $0.950$ \\ 
$-0.025 $ &$0.925$ & $0.944$ & $0.950$ & $0.960$ & $0.958$ & $0.948$ & $0.949$ & $0.948$ & $0.947$ & $0.941$ & $0.948$ & $0.945$ \\ 
$ 0.035 $ &$0.916$ & $0.940$ & $0.929$ & $0.949$ & $0.942$ & $0.966$ & $0.949$ & $0.959$ & $0.952$ & $0.954$ & $0.939$ & $0.937$ \\ 
$  0.095 $ &$0.934$ & $0.951$ & $0.934$ & $0.949$ & $0.946$ & $0.942$ & $0.943$ & $0.943$ & $0.944$ & $0.948$ & $0.944$ & $0.947$ \\ 
$0.155   $ &$0.933$ & $0.952$ & $0.942$ & $0.946$ & $0.936$ & $0.948$ & $0.944$ & $0.952$ & $0.942$ & $0.946$ & $0.948$ & $0.942$ \\ 
$ 0.215 $ &$0.927$ & $0.958$ & $0.927$ & $0.951$ & $0.932$ & $0.941$ & $0.934$ & $0.942$ & $0.953$ & $0.954$ & $0.951$ & $0.939$ \\ 
$0.275 $ &$0.893$ & $0.955$ & $0.913$ & $0.955$ & $0.905$ & $0.951$ & $0.914$ & $0.935$ & $0.926$ & $0.942$ & $0.938$ & $0.949$ \\ 
\hline \\[-1.8ex] 
\multicolumn{12}{l}{meth means we applied the bandwidth selection method, fixed means we used bandwidth $n^{-1/(2J + 1)}$} where J is the kernel order.\\
\end{tabular} 
\end{scriptsize}
\end{table}

\begin{table}[!htbp] \centering 
  \caption{coverage for true parameter, kernel is order 10} 
  \label{} 
  \begin{scriptsize}
\begin{tabular}{@{\extracolsep{5pt}} cccccc}
\hspace{2cm} n = 1000  \hspace{.25cm} &  \hspace{.3cm} n = 2500 \hspace{.25cm} &  \hspace{.3cm} n = 5000 \hspace{.25cm} & \hspace{.3cm} n = 10000\hspace{.25cm}  
& \hspace{.3cm} n = 25000 \hspace{.25cm}  &  \hspace{1cm} n = 50000 \\ 
\end{tabular}
\begin{tabular}{@{\extracolsep{5pt}} ccccccccccccc} 
blip & meth & fixed & meth &   fixed & meth  &  fixed  & meth &  fixed
& meth &  fixed & meth & fixed \\ 
\hline \\[-1.8ex] 
$ -0.145$ &$0.671$ & $0.001$ & $0.436$ & $0$ & $0.298$ & $0$ & $0.294$ & $0$ & $0.338$ & $0$ & $0.325$ & $0$ \\ 
$-0.085 $ &$0.612$ & $0.136$ & $0.593$ & $0.024$ & $0.743$ & $0.001$ & $0.836$ & $0$ & $0.870$ & $0$ & $0.878$ & $0$ \\ 
$-0.025 $ &$0.770$ & $0.166$ & $0.615$ & $0.019$ & $0.405$ & $0.001$ & $0.207$ & $0$ & $0.076$ & $0$ & $0.032$ & $0$ \\ 
$ 0.035$ &$0.850$ & $0.071$ & $0.924$ & $0.001$ & $0.927$ & $0$ & $0.938$ & $0$ & $0.903$ & $0$ & $0.830$ & $0$ \\ 
$0.095 $ &$0.747$ & $0.070$ & $0.895$ & $0$ & $0.912$ & $0$ & $0.924$ & $0$ & $0.906$ & $0$ & $0.801$ & $0$ \\ 
$ 0.155$ &$0.750$ & $0.251$ & $0.859$ & $0.020$ & $0.903$ & $0$ & $0.907$ & $0$ & $0.911$ & $0$ & $0.945$ & $0$ \\ 
$0.215 $ &$0.695$ & $0.947$ & $0.717$ & $0.861$ & $0.793$ & $0.692$ & $0.855$ & $0.370$ & $0.867$ & $0.009$ & $0.877$ & $0$ \\ 
$ 0.275$ &$0.858$ & $0.008$ & $0.817$ & $0$ & $0.707$ & $0$ & $0.493$ & $0$ & $0.147$ & $0$ & $0.010$ & $0$ \\ 
\hline \\[-1.8ex] 
\multicolumn{12}{l}{meth means we applied the bandwidth selection method, fixed means we used bandwidth $n^{-1/(2J + 1)}$} where J is the kernel order.\\

\end{tabular} 
\end{scriptsize}
\end{table} 
\FloatBarrier
\begin{figure}[H]
\caption{}
\includegraphics[width=0.4\linewidth]{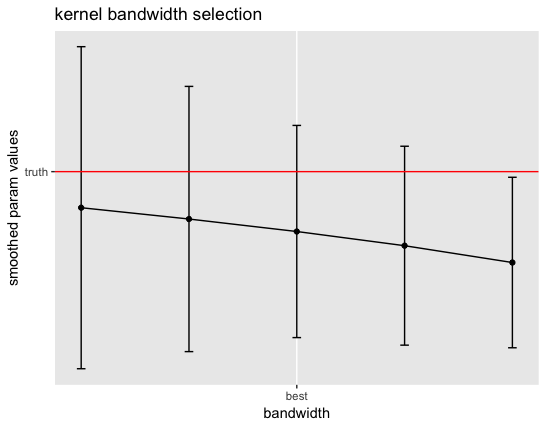}
\end{figure}

It is important that our bandwidth selector allows us to still cover the smoothed parameter and we see by the tables above, that if we apply the selector for kernel order 10, we greatly assist in covering the true parameter and maintain nominal or very near nominal coverage of the smoothed parameter.  Similar results held for lesser order kernels as well.  

\subsection{Simulations for Mispecified Models}
We call these simulations "mispecified" because we use the highly adaptive lasso or HAL \parencite{Laan:2015ab} to recover the model without any specification on functional forms.  The data generating system consisted of the following functions in the order listed.  $W$ is a random normal, $Pr(A=1 \mid W) = g(A \mid W) = expit(-.1-.5*sin(W) - .4*(\vert W \vert >1)*W^2)$ and $E[Y \mid A,W] = expit(.3*A + 5*A*sin(W)^2 - A*cos(W))$.  We simulated 1100 draws from the above data generating system and computed simultaneous TMLE's for the blip values -0.098, -0.018  0.062,  0.142,  0.222,  0.302,  0.382 and  0.462 using bandwidth $2500^{-0.2}$ and an order 1 polynomial kernel.  Similar results held for the uniform kernel. \\

Here we show the huge advantage of data adaptive estimation in obtaining the initial estimates for TMLE\_hal, using the highly adaptive lasso.  HAL will achieve $n^{-0.25}$ $L^2$ rates of convergence to the true outcome regression and propensity score model, assuming the truth is of bounded sectional variation norm and CADLAG (continuous from the right and left hand limits) \parencite{Laan:2015ab}.  TMLE\_glm used initial estimates for the propensity score and outcome regression using logistic regression with main terms and interactions.  We can see it is catastrophic to do so here while HAL recovers very close to nominal coverage and has very little bias.  We can see that TMLE helped remove bias from the HAL initial estimates as well.  The results are displayed in Table 2 and Figure 2 below.  

\begin{table}[!htbp] \centering 
  \caption{TMLE with HAL initial estimates vs glm} 
  \label{} 
    \begin{scriptsize}
\begin{tabular}{@{\extracolsep{5pt}} ccccc} 
\\[-1.8ex]\hline 
\hline \\[-1.8ex] 
 & MSE TMLE\_hal & MSE TMLE\_glm & coverage TMLE\_hal & coverage TMLE\_glm \\ 
\hline \\[-1.8ex] 
blip = -0.145 & $0.00083$ & $0.02003$ & $0.91727$ & $0$ \\ 
blip = -0.085 & $0.00089$ & $0.01683$ & $0.92545$ & $0.01818$ \\ 
blip = -0.025 & $0.00087$ & $0.00528$ & $0.93727$ & $0.58455$ \\ 
blip = 0.035 & $0.00071$ & $0.00373$ & $0.94909$ & $0.81000$ \\ 
blip = 0.095 & $0.00061$ & $0.02173$ & $0.96182$ & $0.10455$ \\ 
blip = 0.155 & $0.00065$ & $0.04723$ & $0.95182$ & $0$ \\ 
blip = 0.215 & $0.00069$ & $0.05803$ & $0.94455$ & $0$ \\ 
blip = 0.275 & $0.00067$ & $0.04528$ & $0.94091$ & $0$ \\ 
\hline \\[-1.8ex] 
\multicolumn{5}{l}{simultaneous TMLE\_hal coverage was 90\%, TMLE\_glm coverage was 3\%}
\end{tabular} 
\end{scriptsize}
\end{table} 

\begin{figure}[H]
  \centering
  \caption{}
  \includegraphics[scale=.09]{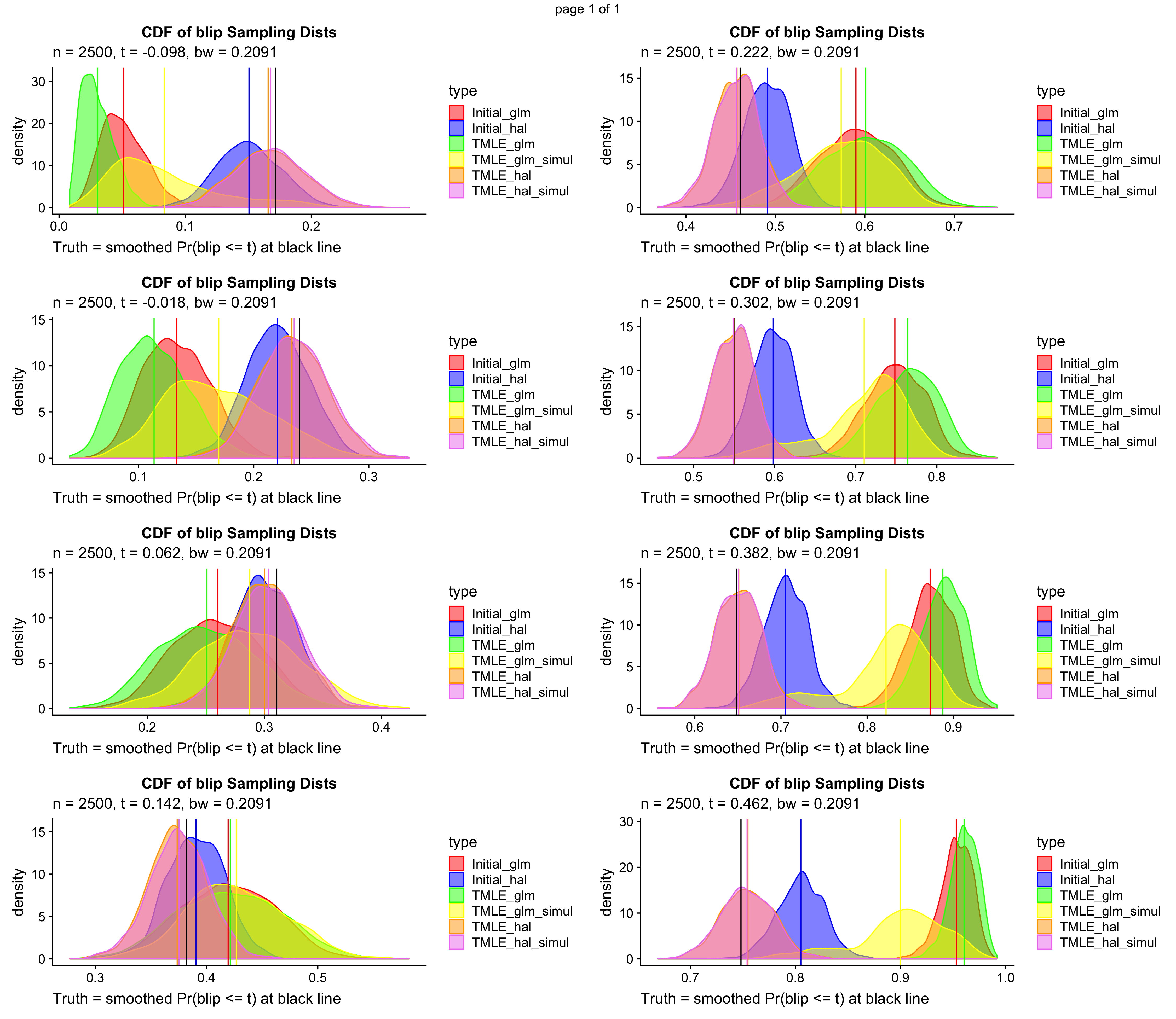}
\end{figure}

\section{Conclusion}
We can see we have developed an estimator with potential to efficiently estimate the kernel smoothed version of the CDF of the blip. We have shown such hinges on data adaptive estimation, in this case with the use of the highly adaptive lasso, to make our initial estimates in the targeted learning \parencite{Laan:2011aa} framework.  Then the targeting helps eliminate bias and provides us with an avenue for immediate inference via the sample standard deviation of the efficient influence curve approximation.  We have shown, in the basic case of one-dimensional $W$ and well-specified models, choosing the bandwidth of optimal order $n^(-\frac{1}{2J+1})$ (see section 10.1.1 with $d=1$) provides normal and unbiased sampling distributions for the smoothed parameter.  The next step is to develop a way to optimally (smallest MSE possible) select the bandwidth and kernel so that the estimator minus the truth blown up by $\sqrt{n\delta}$ is normally distributed.  What we have shown in this paper is a first step but our method of determining monotonicity of the parameter as bandwidth vanishes is somewhat arbitrary.  It also remains to be seen how monotonicity generally holds for small bandwidths.  For instance, if the monotonicity changes direction for a small bandwidth, our proposed selector might be problematic.  

\newpage
\printbibliography

 \end{document}